\documentstyle[12pt]{article}
\begin{document}
\centerline{\large{\bf The Penna Model for Biological Ageing on a Lattice:}} 
\centerline{\large{\bf Spatial Consequences of Child-Care}}  
\bigskip

\centerline{\large{A.O. Sousa and S. Moss de Oliveira}}

\bigskip
\centerline{\it Instituto de F\'{\i}sica, Universidade Federal Fluminense}

\centerline{\it Av. Litor\^anea s/n, Boa Viagem, Niter\'oi 24210-340, RJ, 
Brazil}


\bigskip
\noindent{\bf Abstract}
\bigskip

We introduce a square lattice into the Penna bit-string model for biological 
ageing and study the 
evolution of the spatial distribution of   
the population considering different strategies of  
child-care. Two of the strategies are related to  
the movements of a whole family on the lattice: in one case the mother cannot 
move if she has any child younger than a given age, and in the other case if 
she moves, she brings these young children with her. A stronger 
condition has also been added to the second case, considering that young 
children die with a higher 
probability if their mothers die, this probability decreasing with age.  
We show that a 
highly non uniform occupation can be obtained when child-care is considered, even 
for an uniform initial occupation per site.  
We also compare the standard survival rate of the model with that obtained when 
the spacial lattice is considered (without any kind of child-care).
\section{Introduction}

The Penna bit-string model for biological ageing 
was published in 1995 [1], and  since then it has been   
increasingly used to study different characteristics of  
real populations, as for instance the inheritance of longevity [2] and the 
advantages of sexual reproduction [3] (for a  
review, see [4,5] and references therein). 
Parental care was first introduced in the model by Thoms et al 
[6]. More recently, it has been shown that different strategies 
of child-care can lead to smaller or higher effects on the survival 
probabilities of the population [7], and that when combined with reproduction 
risk, it leads to a self-organisation of female menopause in sexual populations 
[8]. 

In this paper we are mainly interested in two points: 
the first one is to study the spatial distribution of  
populations subjected to different strategies of child-care. 
Two of our strategies are  
specifically related to the movements of the mother on the lattice, and so 
different from those mentioned above.
The third one is also related to the survival of the child, depending if the mother 
is alive or not [7,8,9]. The second point is to compare the usual survival rate 
of the model with that obtained when the lattice is introduced.
 
In the next section we describe how the asexual Penna model works 
on a lattice and our strategies of child-care. In section 3 we present the  
results considering child-care and in section 4 the results when the standard 
model is compared with the lattice one (without any child-care strategy). 
In section 5 we present our conclusions.
    
\section{The Penna Model on a lattice and strategies of child-care}

In the asexual version of the Penna model each individual is represented by a 
single bitstring of 32 bits, that plays the role of a chronological genome.
Each individual can live at most for 32 timesteps (``years''). 
The first bit gives the information if the individual will suffer 
from the effects 
of a genetic disease in its first year of life. The second bit corresponds 
to the second year of life, and so on. Diseases are represented by bits 1. 
Whenever a bit 1 appears it is computed 
and the total amount of diseases until the current individual's age is compared 
with a 
limit $T$. The individual dies when the limit is reached. There is a minimum 
reproduction age $R$, from which the individual generates  
with probability 0.5 (in this paper), $b$ offspring every year. We also consider 
a ``pregnancy'' period: after giving birth, a mother stays one timestep (year) 
without reproducing. The  
offspring genome is a copy of the parent's one, with $M$ random mutations. Only 
deleterious mutations are considered: if a bit 0 is randomly chosen in the 
parent's genome, it is mutated to 1 in the offspring one. If a bit 1 is randomly 
chosen, it stays 1 in the offspring genome.  

Besides dying if the number of allowed diseases exceeds $T$ or for reaching 32 
years, an individual may also die for lack of space and food. Without 
a spatial lattice, such lack is taken 
into account through the general Verhulst factor $V=1-N(t)/N_{max}$, where $N(t)$ 
is the 
actual size of the population at time $t$ and $N_{max}$ is the maximum size 
(carrying capacity) the population can sustain, which is defined at the beginning 
of the simulation [1]. At every timestep and for each individual, a random number 
between 0 and 1 is generated and compared with $V$. If this random number is 
greater than $V$, the individual dies independently of its age or genome.   

In the present case each individual lives in a given site (i,j) of a square 
lattice. Instead of considering a single carrying capacity for the whole 
population, {\it we consider a maximum allowed occupation per site}. Moreover, 
at every timestep 
each individual {\it has a probability} to walk to the neighbouring site that 
presents the smallest occupation, if this occupation is also smaller or equal to 
that of the current individual's site. The Verhulst factor is now given by: 
$$V_{i,j}=1-N_{i,j}(t)/N_{i,j(max)} \, \, .$$ 
In this way, it is possible to consider the same carrying capacity for all sites, 
or to define regions with more or less resources. 
Each individual is first tested: 
if it does not die because of genetic diseases, lack of food and space (Verhulst) 
or for reaching 33 years, it moves according to the probability mentioned 
above. 
We start the simulations 
randomly distributing one individual per site on a diluted lattice. That is, 
if an already occupied site is chosen for a new individual, the choice is 
disregarded and another random number is generated.  

Our strategies of child-care which are purely related to the movements on the 
lattice consist 
in defining a period $C_m$ during which the mother and the child are forced to stay 
together: no 
child can move alone before reaching an age greater than $C_m$. In this way, 
$C_m$ {\it is the period during which the child is under maternal care}.   
We have studied the following cases: 
(a) if the  
mother moves, she brings all these young children with her; 
(b) the mother cannot move if she has any child still under maternal care.

We also considered a condition, that can be added to any of the other 
two or even be considered alone: it consists in killing the motherless child with  
age $\le C_m$ according to a given probability which decreases with age.
We have considered $C_m =2$, a probability 0.9 that the child dies in its first 
year of life if its mother dies, and a probability 0.3 if it happens in its second 
year of life. This strategy is in fact an improved version of that used before in 
refs. [7,8], where the offspring die if the mother dies, with probability 1, 
at any age inside the child-care period. This improvement is based on data given 
in ref. [9], fig.5, about baboons and lions. 
In next section we present our results for cases (a), (b) and (c), where 
(c) corresponds to case (a) plus the survival condition just mentioned. 

\section{Results comparing child-care strategies}     

In this paper we have worked with the same maximum occupation per site. The 
figures presented in this section correspond to the following parameters:
\bigskip

\noindent {\bf I -} General parameters of the model

\noindent 1) Initial population = 10,000 individuals;

\noindent 2) Limit number of diseas $T$ = 2;

\noindent 3) Carrying capacity (per site) = 34; 

\noindent 4) Minimum reproduction age = 9;

\noindent 5) Birth probability = 0.5; 

\noindent 6) Birth rate $b$ = 2; 

\noindent 7) Total number of steps = 800,000.

\bigskip  

\noindent {\bf II -} Parameters related to the lattice and child-care 

\noindent 1) Lattice size = 150 X 150 sites;

\noindent 2) Probability to walk = 0.2;  

\noindent 3) $C_m = 2 \rightarrow$ age up to which to move alone is forbidden. 

\bigskip
With these parameters a maximum population of 160,000 individuals is 
reached, this number decreasing and stabilizing between 25,000 and 5,000 
individuals, depending on the child-care strategy.

\bigskip
In figure 1 we show the final configuration of the lattice for  
case (a), and in figure 2, for case (b). In figure 3 we 
present the results for case (c), where two strategies of child care are 
working together, as mentioned at the end of section 2.   

From these figures it can be noticed that the final configuration is strongly 
dependent on the child-care strategy. Comparing cases (a) and (b), 
it can be 
noticed that when the mother cannot move, more empty sites are produced, as  
a simple consequence of the Verhulst factor: more individuals accumulate at the 
same site and die for lack of space and food. 

When the condition that the child dies if 
the mother dies is added (fig.3), the final configuration presents more empty sites 
than fig.1. In this third case  
two effects are superimposed: a stronger action of the Verhulst factor (if the 
mother is randomly chosen to die, all young children also die), {\it and deaths 
caused by weak genotypes} (mothers that die for accumulation of deleterious 
mutations). Since the Verhulst factor does not interfere with the genetic 
features of the population, the survival rates of cases (a) and (b) are the 
same, and different from that of case (c). 
In fact, the genetic deaths of mothers with weak genotypes select the best fitted 
families 
to survive, increasing the lifespan [7,8], as can be seen from figure 4.  
On the other side, the total population decreases the more the stronger is the 
strategy of child-care considered [6,7,8]. 

\section{Comparison between the standard results and those obtained with the 
lattice}

A natural question that appears when the lattice is introduced is: ignoring 
any child-care strategy, how to compare the standard results of the Penna model 
with those when the lattice is introduced? Or, which is the behaviour of the 
survival rate with and without the lattice? In order to answer this question 
we performed several simulations and observed that: 

\noindent a) For a probability to walk equal to one, the survival rates 
of the model with and without the lattice are the same, and no 
special care must be taken. It means that 
if one considers, as usual, a given carrying capacity $N_{max}$ for the whole  
population, the same survival rate can be obtained simply considering a 
carrying capacity per site equal to the same $N_{max}$ divided by the number 
of sites. 

\noindent b) As the probability to walk decreases, finite size effects appear. 
They are observed through a corresponding decrease of the maximum lifetime, 
that is, the survival rates drop to zero before the standard ones, as can be  
seen from fig.5a (where the maximum occupation per site is 34). In these cases, 
if one is interested in comparing   
the lattice results with those of the standard model, a higher 
occupation per site 
must be considered. If this occupation is large enough, the results are the 
same, as shown in fig.5b, where the maximium occupation per site is 200. 
A maximum occupation per site equal to 100 produces the same result and is 
so enough to avoid finite size effects.   

\noindent c) Periodic boundary conditions cannot avoid the finite size effect 
mentioned above, for small occupations per site.

\section{Conclusions}

We show how to implement the Penna model on a lattice, in order to study the 
spatial evolution of the population. A walking 
probability is given to each individual, and the Verhulst factor is now per site,  
i.e., there is a maximum capacity per site. We start the simulations randomly 
distributing one individual per site, on a diluted lattice. 
We consider that no individual with age 
$\le C_m$ can move alone, i.e., while under maternal care.    
The following strategies of child care have been considered: (a) If the mother 
moves she brings the offspring under child maternal care with her; (b) The mother 
cannot move 
if she has any child under care. Comparing these two situations, we notice that 
the second one produces more empty sites than the first one, as a consequence of 
the Verhulst factor (the occupations of many sites exceed the maximum allowed 
capacity). An extra condition has also been added to the first strategy: if the 
mother 
dies, the offspring has a probability to die that depends on its age. In this 
case, also more empty sites are produced, if compared to the first case. 
We present the survival rates of the three cases, showing that only the last 
one interferes with the population genetics.

We also compare the survival rates obtained with the standard model with those 
obtained when the lattice is introduced (without any kind of child-care). We show 
that for large enough maximum occupations per site, these survival rates are the 
same.
\bigskip

\noindent Acknowledgements: To P.M.C. de Oliveira, J.S. Sa Martins and A.T. 
Bernardes for 
usefull discussions; to CAPES and CNPq for financial support.
 
\bigskip
\centerline{\bf References}

\begin{description}
\item {[1]} T.J.P. Penna, J.Stat.Phys. {\bf 78} (1995) 1629.
\item {[2]} Paulo Murilo C. de Oliveira, Suzana Moss de Oliveira, Americo T. 
Bernardes and Dietrich Stauffer, Lancet, {\bf 352} (1998) 911.
\item {[3]} J.S. S\'a Martins and S. Moss de Oliveira, Int.J.Mod.Phys. {\bf C9} 
(1998) 421.
\item {[4]} A.T. Bernardes, {\it Annual Reviews of Computational Physics IV}, 
(1996) 359. Edited by D. Stauffer - World Scientific, Singapore. 
\item {[5]} S. Moss de Oliveira, Physica {\bf A257} (1998) 465; Suzana Moss de 
Oliveira, Paulo Murilo Castro de Oliveira and Dietrich Stauffer, {\it Evolution, 
Money, War and Computers}, Teubner, Stuttgart-Leipzig (1999).
\item {[6]} J. Thoms, P. Donahue, D. Hunter and N. Jan, J.Phys. {\bf I5} (1995) 
1689.
\item {[7]} K.M. Fehsenfeld, J.S. S\'a Martins, S. Moss de Oliveira and A.T. 
Bernardes, Int.J.Mod.Phys. {\bf C9} (1998) 935.
\item {[8]} S. Moss de Oliveira, A.T. Bernardes and J.S. S\'a Martins, 
Eur.Phys.J. {\bf B} (1998) in press.
\item {[9]} C. Packer, M. Tatar and A. Collins, Nature {\bf 392} (1998) 807.
\end{description}

\newpage
\centerline{\bf FIGURE CAPTIONS}
\bigskip

\noindent Fig.1 - Final distribution of the population on the lattice considering 
that if the mother moves, she brings the children under maternal care 
($age \le 2$) with her (case a).

\bigskip
\noindent Fig.2 - Final distribution of the lattice considering that the mother 
cannot move if she has children under maternal care (case b).

\bigskip
\noindent Fig.3 - Final distribution of the lattice considering that if the mother 
moves she brings the children under maternal care with her, and that if she dies, 
her children have a probability 0.9 to die in their first year of life, and 
0.3 in their second year of life (case c).

\bigskip
\noindent Fig.4 - Survival rates for the cases: (a)- filled circles; 
(b)- triangles; (c)- squares. 

\bigskip
\noindent Fig.5a - Survival rates of the standard model without lattice 
(triangles), and with lattice for different probabilities to walk.  
In all cases the initial population $N(0)=10,000$   
and the common parameters ($T,b,R$,birth probability) 
are those described in section 3. In the lattice cases, the maximum occupation 
per site is 34 and the lattice size is 150X150. In the standard case 
$N_{max}=10N(0)$.   

\bigskip
\noindent Fig.5b - Triangles: the same standard survival rate of fig.5a; 
Full squares: the
survival rate with lattice for a maximum occupation per site equal to 
200 and a probability to walk equal to 0.2.

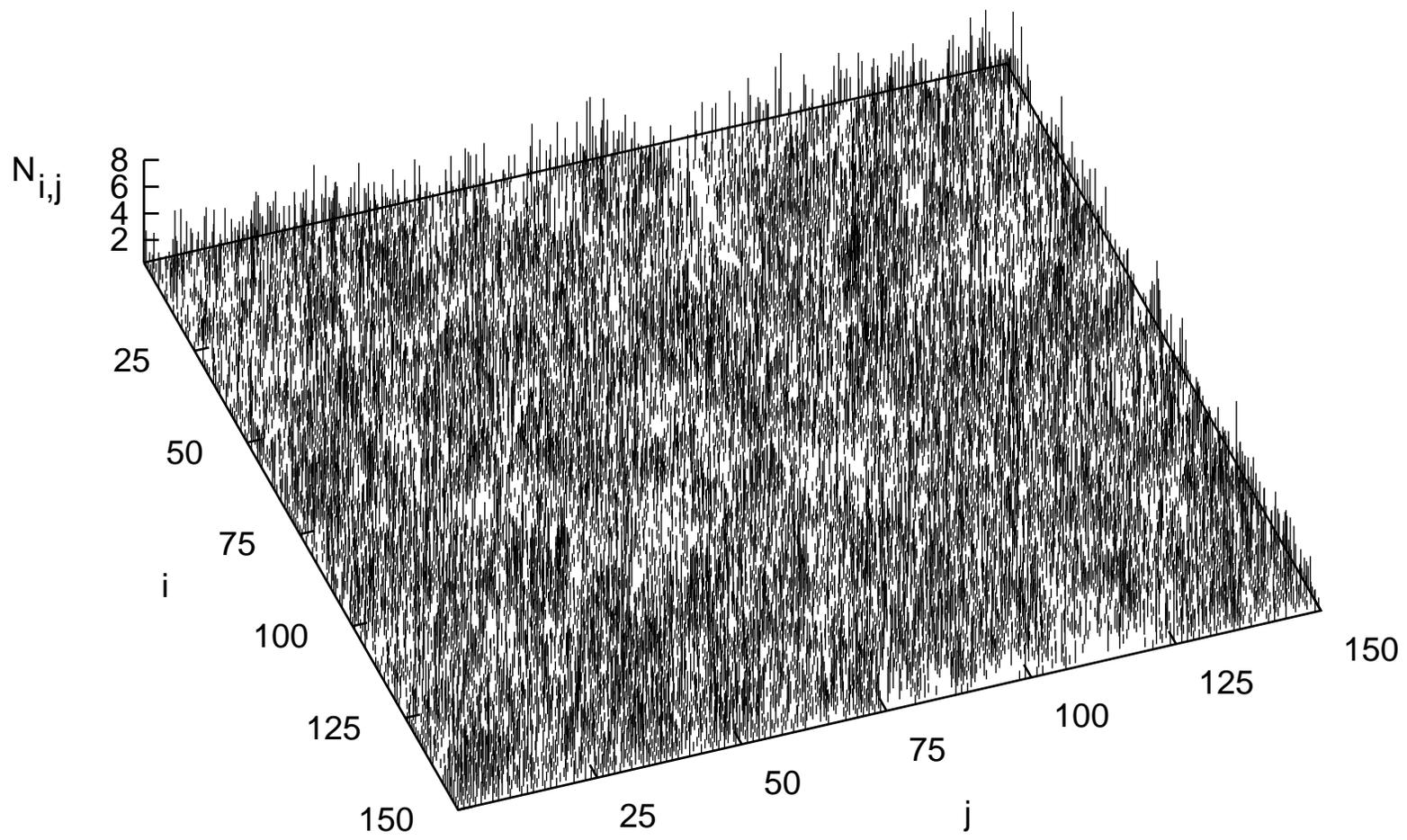
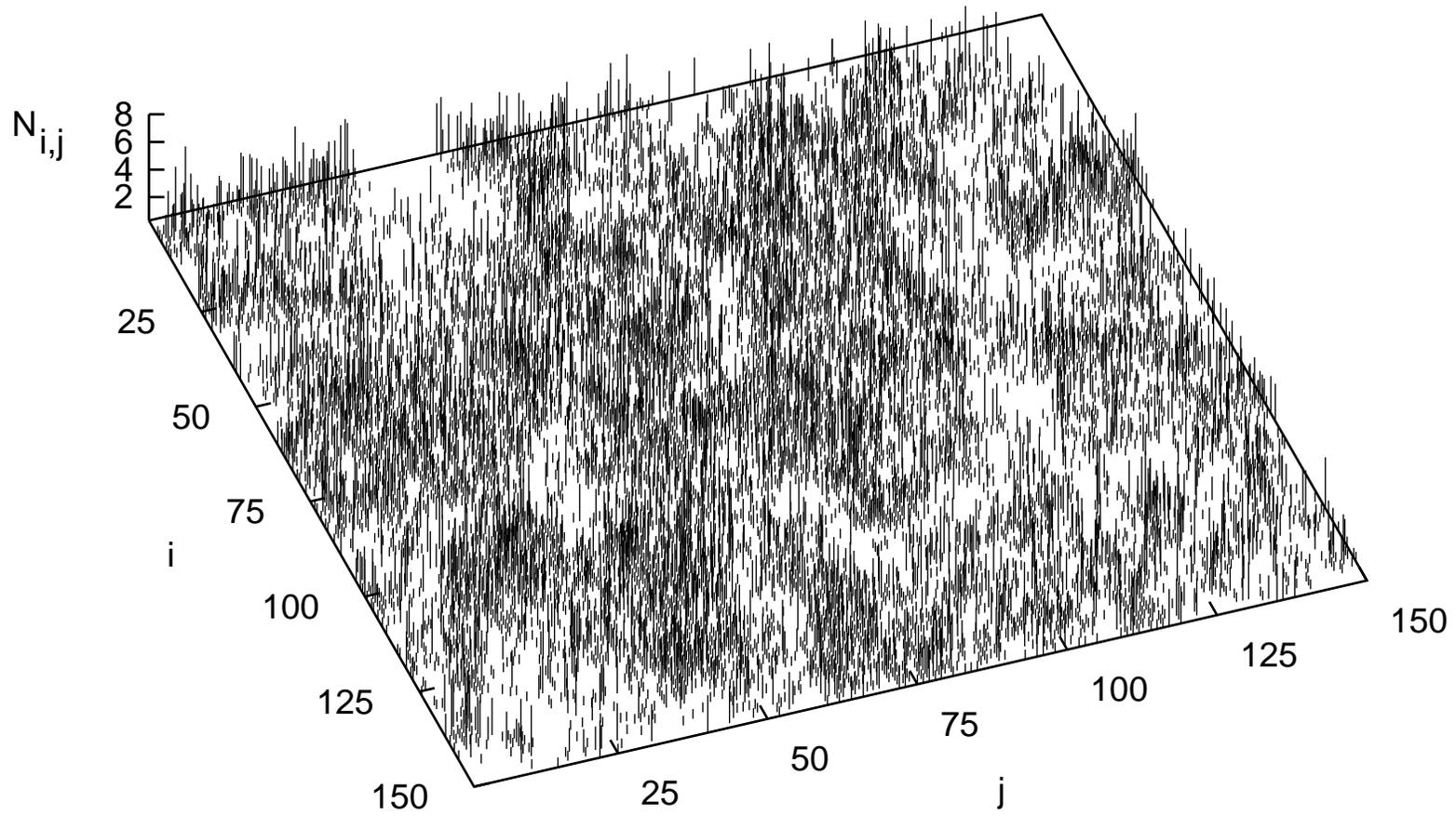
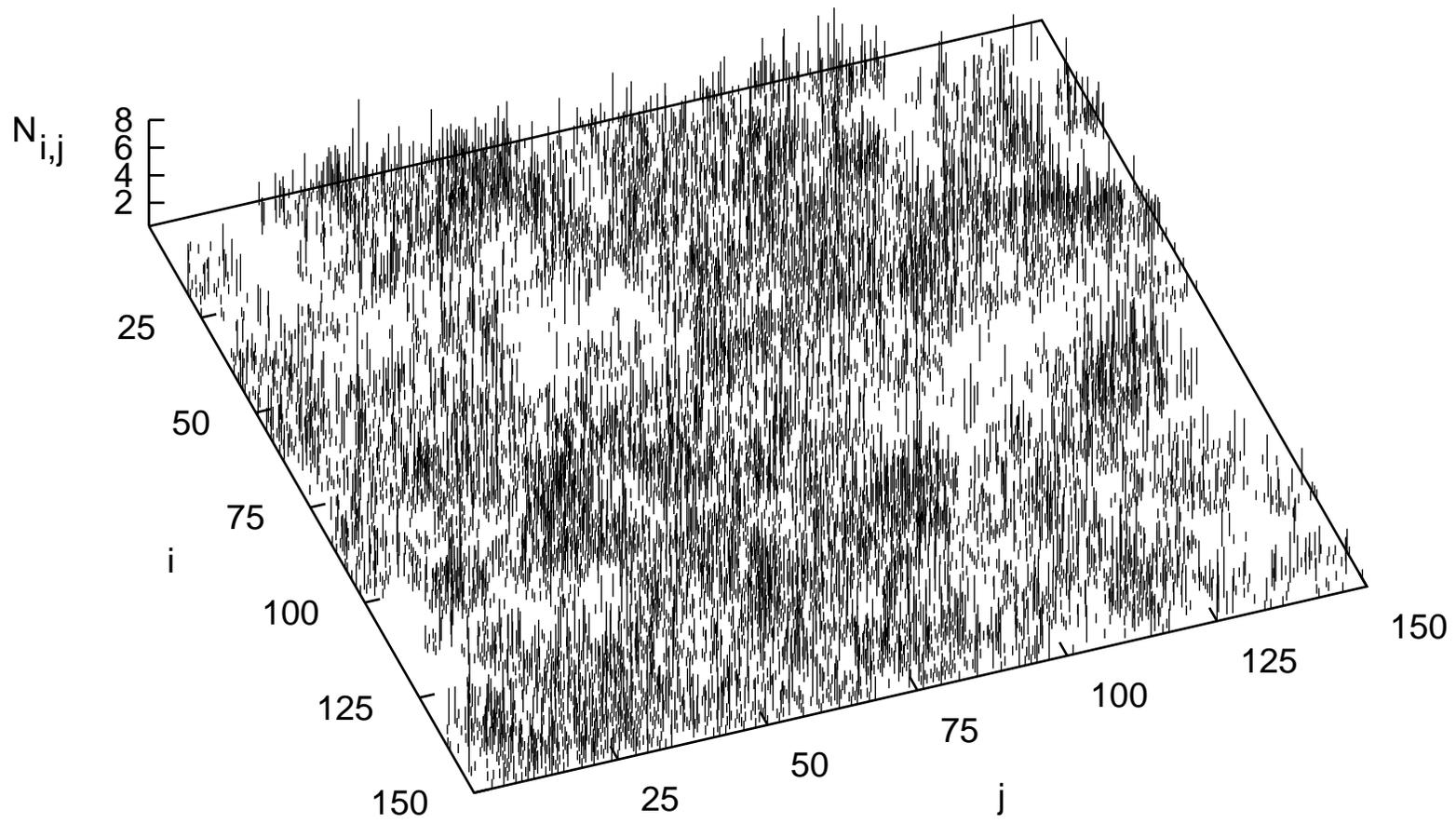
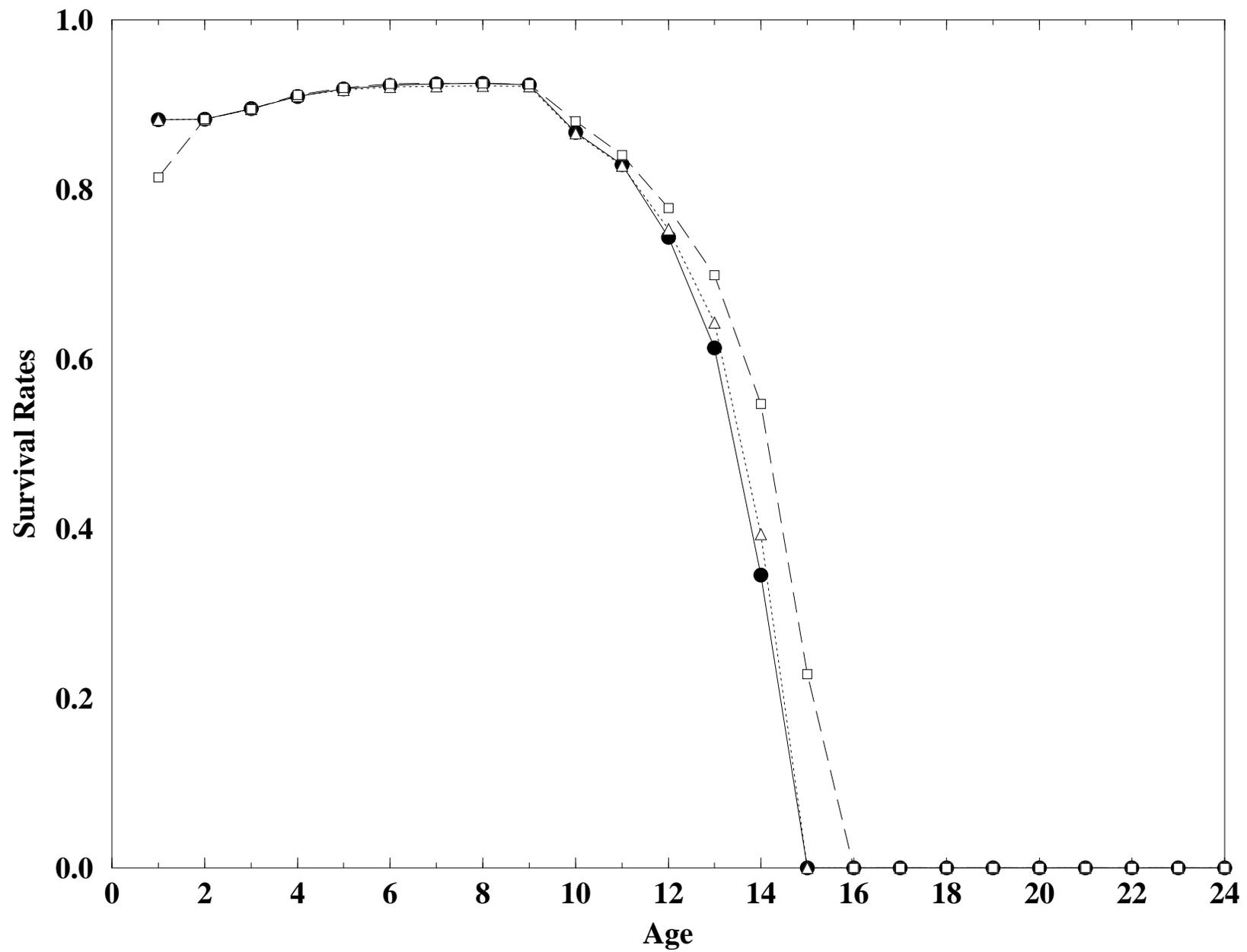
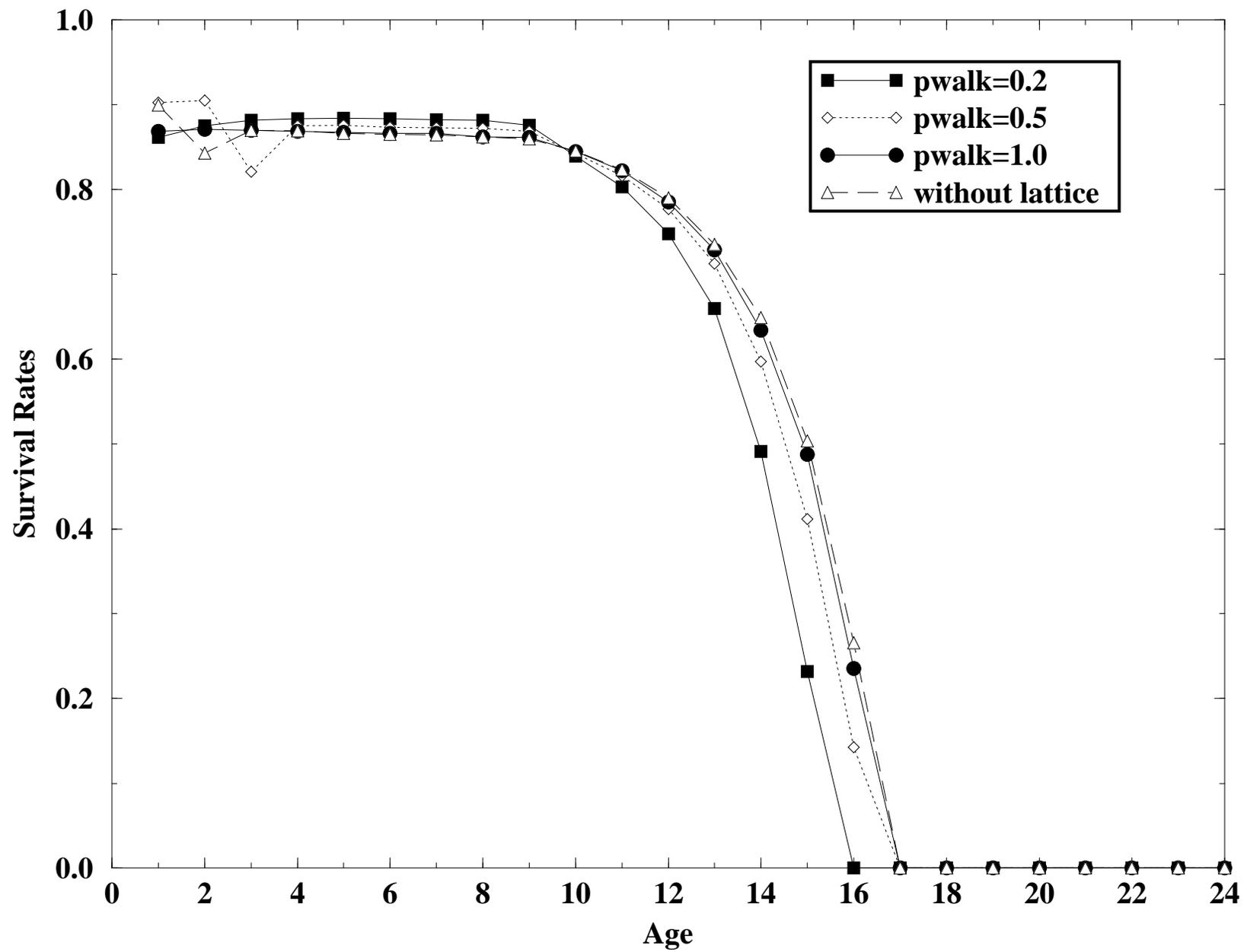
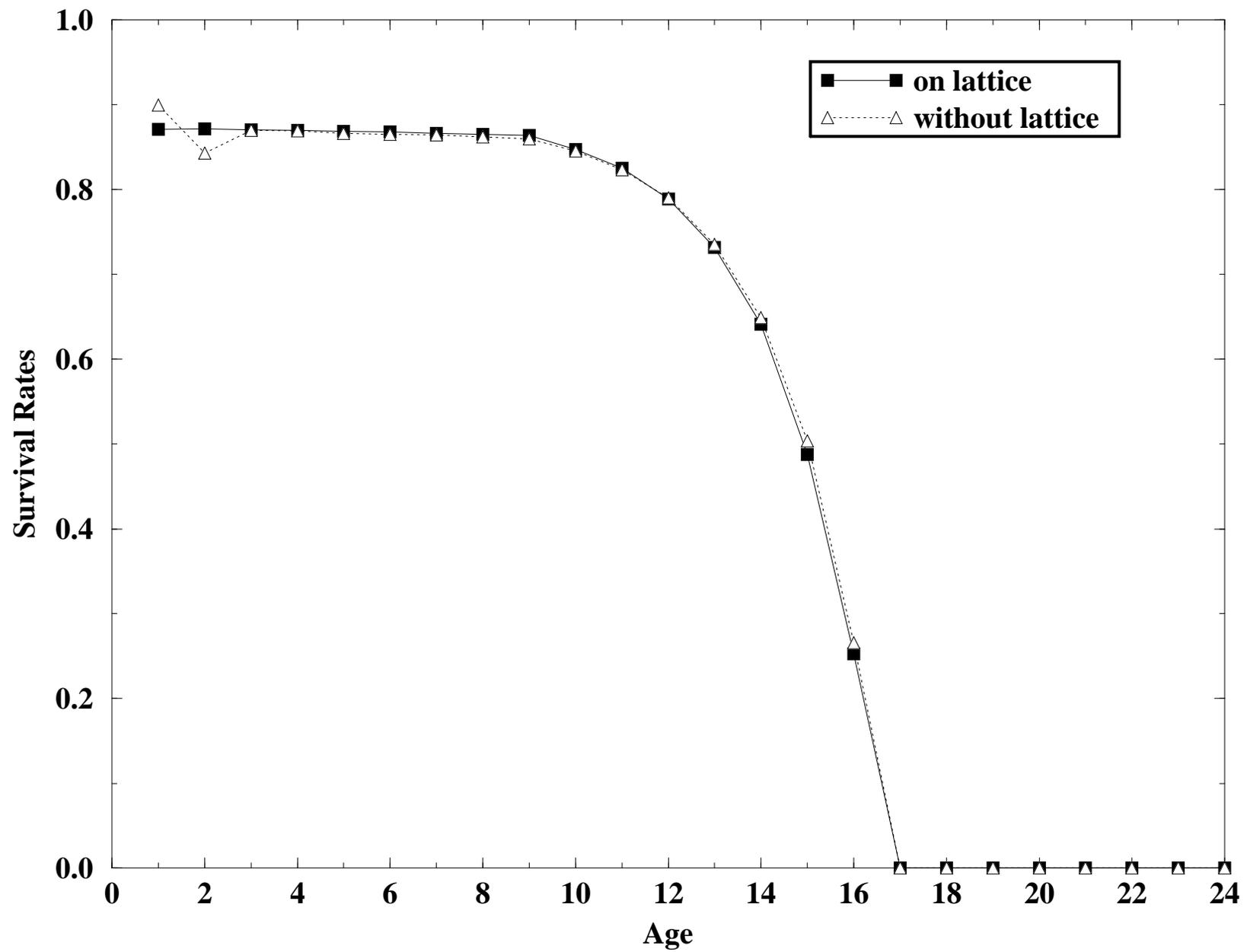
\end{document}